\documentclass[11pt,twoside,letterpaper]{article} %% The same for {book}
\usepackage{times,fancyhdr}
\usepackage[dvips]{graphicx}
\usepackage{amsmath,epsfig,graphicx,amssymb}


\makeatletter
\def\cleardoublepage{\clearpage\if@twoside \ifodd\c@page\else%
    \hbox{}%
    \thispagestyle{empty}%
    \newpage%
    \if@twocolumn\hbox{}\newpage\fi\fi\fi}
\makeatother

\usepackage{graphicx}
\usepackage{adjustbox}
\usepackage{tikz}
\usetikzlibrary{arrows,shadows,backgrounds,calc,decorations.pathreplacing,decorations.pathmorphing,positioning}
\usepackage{pgfplots}
\usepackage[toc,acronym]{glossaries}
\pgfplotsset{compat=1.13}
\definecolor{webgreen}{rgb}{0,.5,0}
\definecolor{webbrown}{rgb}{.6,0,0}
\definecolor{webyellow}{rgb}{0.98,0.92,0.73}
\definecolor{webgray}{rgb}{.753,.753,.753}
\definecolor{webblue}{rgb}{0,0,.8}
\definecolor{webgreen}{rgb}{0, 0.5, 0} % less intense green
\definecolor{webred}{rgb}{0.5, 0, 0}   % less

\begin{document}
	  \newacronym{HPL}{HPL}{High Performance Linpack}
	\newacronym{HPCG}{HPCG}{High Performance Conjugate Gradients}
	\newacronym{SPA}{SPA}{Single Processor Approach}

\title{
{\begin{flushleft}
\vskip 0.45in
{\normalsize\bfseries\textit{Chapter~1}}
\end{flushleft}
\vskip 0.45in
\bfseries\scshape How deep the machine learning can be\footnote{Submitted to appear in book "A closer look at deep learning", by Nova}}}
\author{\bfseries\itshape J\'anos V\'egh
	\thanks{E-mail address: Vegh.Janos@gmail.com
		% and Alin.Tisan@rhul.ac.uk
	}\\
Kalim\'anos BT, Debrecen, Hungary
}
\date{}
\maketitle
\thispagestyle{empty}
\setcounter{page}{141}
% ------- [First Page Running Head] - place it immediately after title! ------
\thispagestyle{fancy}
\fancyhead{}
\fancyhead[L]{In: Book Title \\
Editor: Editor Name, pp. {\thepage-\pageref{lastpage-01}}} % needs \label{lastpage-01} on the last page.
\fancyhead[R]{ISBN 0000000000  \\
\copyright~2006 Nova Science Publishers, Inc.}
\fancyfoot{}
\renewcommand{\headrulewidth}{0pt}
%------------------------------------------------------------------------------

\begin{abstract}
Today we live in the age of artificial intelligence and machine learning; from small startups to HW or SW giants, everyone wants
to build machine intelligence chips, applications.
The task, however, is hard: not only because of the size of the problem: the technology one can utilize  (and the paradigm it is based upon) strongly degrades the chances to succeed efficiently.
Today the single-processor performance practically reached the limits
the laws of nature enable. The only feasible way to achieve the needed high computing performance seems to be parallelizing many sequentially working units. The laws of the (massively)  parallelized computing, however, are different from those experienced in connection with assembling and utilizing systems comprising just-a-few single processors. As machine learning is mostly based on the conventional computing (processors), we scrutinize the (known, but somewhat faded) laws of the parallel computing, concerning AI.
This paper attempts to review some of the caveats,
especially concerning \textit{scaling} the computing performance of the AI solutions.
\end{abstract}

% Note: When having missing abstracts, please let the 2in hole:
\vspace{2in}

%\noindent \textbf{PACS} ?;?05.45-a, 52.35.Mw, 96.50.Fm.
%\vspace{.08in} \noindent \textbf{Keywords:} ???Energy of vacuum state.

%% Other situations:
\vspace{.08in}% \noindent {\textbf AMS Subject Classification:} 53D, 37C, 65P.

% ------------ [Running Heads - for odd and even pages] - please insert it only on page 2!
\pagestyle{fancy}
\fancyhead{}
\fancyhead[EC]{J\'anos V\'egh
	% and Alin Tisan
}
\fancyhead[EL,OR]{\thepage}
\fancyhead[OC]{How deep machine learning can be}
\fancyfoot{}
\renewcommand\headrulewidth{0.5pt}
%------------------------------------------------------------------------------
%\tableofcontents

\section{Introduction}
The computer initially
had the task to automate lengthy computations on a small amount of data, there was only one processor and the memory access time was in the same order of magnitude as the time of performing a machine instruction.
At that time, the limiting resource was the speed the computer could perform the needed elementary operations with.
For today the development of the technology and the way of utilizing computers for AI has changed to the exact opposite: a huge (and growing) number of processors shall process a huge amount of data. The processors spend a considerable fraction
of their working time with waiting for the availability of the data (either from the instruction or data memory), or shared resources (like the high-speed bus), or communicating the results of the computations from/to the fellow processors, or are just idle because of the computing paradigm and its implementation.
In addition to the parallel performance issues detailed below, similar issues occur under other specific extreme conditions too, but they receive much less spotlight. Some of these examples are:
\begin{itemize}
	\item the very inflexible and inefficient separation of the hardware \textit{HW} and software \textit{SW} in a real-time multitasking system leads to the phenomenon known
	as 'priority inversion'~\cite{PriorityInversion:1993}
	\item the massive  inflexible architectures suffer from frequent component errors
	\item the complex cyber-physical systems
	must be equipped with excessive computing facilities to provide real-time operation
	\item a serious challenge is to deliver the vast amount of data from the big storage centers to
	the places where the processing capacity is concentrated
	\item only a fragment of the continuously running cores can be utilized ~\cite{Computing_Dark_Silicon_2017} simultaneously
	\item supplying  energy to the huge number of computers has begun more and more problematic
\end{itemize}
All the issues have a common reason: the \textit{classic computing paradigm that reflects
	the state of the art of 70 years ago}~\cite{SoOS:2010}.
Computing needs renewal~\textbf{\cite{RenewingComputingVegh:2018}}.

In section~\ref{sec:milestones}, we briefly review
some of the terms, milestones, and issues of the parallelization.
Section~\ref{sec:ParallelEfficacy} introduces the terms of the efficiency of parallelization and demonstrates that parallelization introduces a new performance limit (on the top of the already known limitations of the single-processor performance).
Section~\ref{sec:ModernComputing} presents that when using computing under extreme conditions
(such as the vast number of cores in the
parallelized systems or extremely high communication to computing ratio) unusual phenomena occur. The
section draws analogies with the classic versus modern physics:
the new (and unexplained) phenomena experienced under extreme conditions
forced revising our basic knowledge.

As in all parallelized systems, the different synchronization mechanisms play a unique role also in AI,  as detailed in section~\ref{sec:synchronizationAI}. This section comprises
a case study about brain simulation: at those extreme large scales of parallelized systems the operating principle and implementation
make the 'quantal nature' of computing time dominating
factor of performance loss.

In the light of the previous sections,   section~\ref{sec:efficacyAI} discusses how the present architectural principles define the resulting efficacy.

\section{Major terms and milestones of parallelization\label{sec:milestones}}
It was early discovered that the
age of conventional architectures was over~\cite{AmdahlSingleProcessor67,GodfreyArchitecture1986}; the only question remained open for decades later whether
the "game is over", too~\cite{ComputingPerformanceBook:2011}.

\subsection{The moon-shot of linear performance scaling}
Despite the early warnings~\cite{AmdahlSingleProcessor67,ScalingParallel:1993} the computing followed the path dictated by the Moore's Law:
the moon-shot of unlimited growing of computer performance, not considering that the growth seems to be exponential \textit{only} at the beginning~\cite{ExponentialLawsComputing:2017} and
"\textit{a trend that can't go on ad infinitum.}"\cite{ExascaleGrandfatherHPC:2019}.
The development efforts concentrated on achieving high-performance single-processor, but
there are scientific reasons why the 'single-processor' computation shall reach a limit~\cite{NeuromorphicComputing:2015,LimitsOfLimits2014}.
Another moon-shot today is parallelization, the only way to
increase the computing performance of the system with conventional processors is to parallelize the sequentially working processors.
Even today an (exponentially) linear dependence is expected~\cite{ChinaExascale:2018} and the "gold rush" for
achieving exascale performance~\cite{ScienceExascaleRace:2010} is going on.
Even in the most prestigious journals~\cite{StrechingSupercomputers:2017,Scienceexascale:2018}.

However, as demonstrated decades ago, \textit{the computing performance not only does not increase linearly with the number of computing units, but also decreases after exceeding a critical number of processors}~\cite{ScalingParallel:1993}. Although the EFlops (payload) performance has not yet been achieved, already the $10^4$ times higher supercomputer
performance is planned~\cite{ChinaExascale:2018}.
It looks like that in the feasibility studies  an analysis,
whether an inherent performance bound exists, remained out of sight in USA~\cite{Scienceexascale:2018,NSA_DOE_HPC_Report_2016} or in  EU~\cite{EUActionPlan:2016}
or in Japan~\cite{JapanExascale:2018} or in China~\cite{ChinaExascale:2018}.
However, at extreme large scale systems serious performance limitations occur~\textbf{\cite{VeghPerformanceWall:2019,VeghBrainAmdahl:2019}}. The parallelized sequential processing has different rules of game~\cite{ScalingParallel:1993},~\textbf{\cite{VeghModernParadigm:2019}}: the performance gain ("the speedup") has its inherent bounds~\textbf{\cite{VeghRoofline:2019}}. It has recently been admitted that it "\textit{can be seen in our current situation where the historical ten-year cadence between the attainment of megaflops, teraflops, and petaflops has not been the case for exaflops}"\cite{ExascaleGrandfatherHPC:2019}.

\begin{figure}
	\hskip-1cm	\maxsizebox{1.1\textwidth}{!}
	{
		\includegraphics[scale=.95]{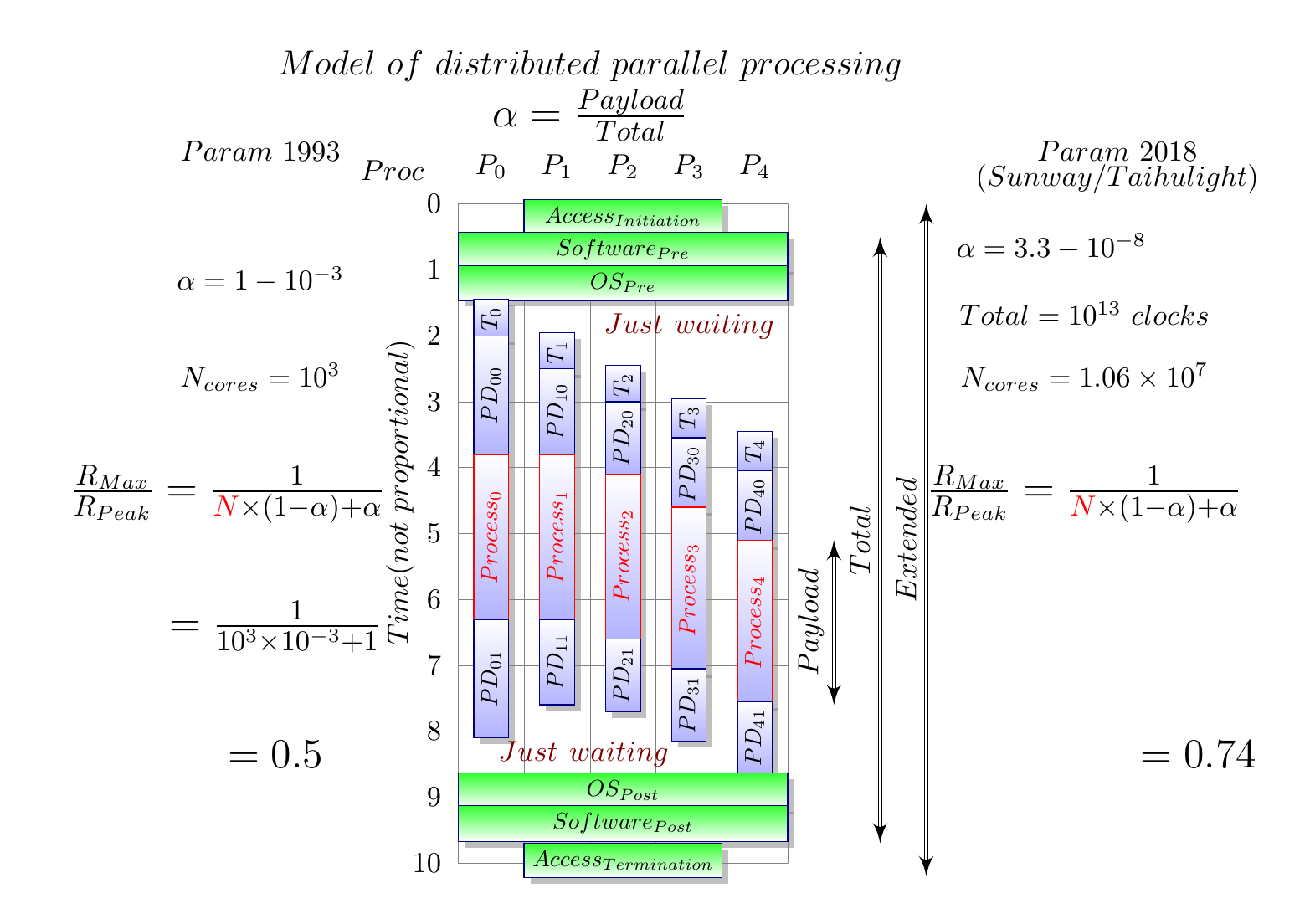}
	
	}
	\caption{The simple model of distributed parallel processing. With proper interpretation of the terms, it can describe all kinds of distributed parallel processing.\label	{fig:AmdahlModelAnnotated}}
\end{figure}

\subsection{Model of parallelized sequential processing}
The only form of cooperation of processing units that researchers elaborated till now, was parallelizing the operation of some otherwise sequentially working processors. However, there are attempts to transfer the workload from one core to another~\cite{Congy:CoreSpilling:2007,ARM:big.LITTLE:2011}; the immediate register-to-register transfer~\cite{CooperativeComputing2015} enabled to bo build supercomputer with 10M cores and considerably increased the efficiency of the benchmark HPCG~\cite{TaihulightHPCG:2018}.
As it was predicted~\cite{AmdahlSingleProcessor67},
the principle of parallel processing introduces its own limitations on the resulting parallel performance.
The simple model shown in Fig.~1 %\ref{fig:AmdahlModelAnnotated}
may help to understand the parallelized sequential operation.
Despite its non-technical nature and simplicity,
the model correctly describes the development of supercomputing
in its quarter of century history and predicts the present stalling and limitations.

The supercomputing is a distributed parallelized computing: one of the processors distributes the job
for the rest of the processors that perform their parts sequentially.
A one-time initialization (as well as termination) is needed both in the SW and the OS. Then the first processing unit must spend some $T_n$ time with addressing the processor (also part of the OS). After this, the propagation delay takes place ($PD_{n0}$ to and $PD_{n1}$ from the $n$th processor)\footnote{Notice that the $T_n$ and $PD_{nx}$ terms can be combined to achieve lower total execution time, see Fig.~1.}. The same timing follows after that the computation is performed by the corresponding processing unit.
Notice that the first processing unit must work alone until
all tasks scheduled to the fellow processors, and also it must wait until the last fellow finishes its task.

Amdahl's original intention was to call the attention to that
parallelizing  sequentially processing units
introduces serious performance limitations.
His successors formulated  Amdahl's law  as
\vspace{-.3\baselineskip}
\begin{equation}
S^{-1}=(1-\alpha) +\alpha/N \label{eq:AmdahlBase}
\end{equation}

\noindent where $N$ is the number of parallelized code fragments,
$\alpha$ is the ratio of the parallelizable portion to the total,
$S$ is a measurable speedup.
Although calculating $\alpha$ for the today's sophisticated hardware/software systems is extremely hard,
one can express the parallel portion as

\vspace{-.3\baselineskip}
\begin{equation}
\alpha = \frac{N}{N-1}\frac{S-1}{S} \label{equ:alphaeff}
\end{equation}

\noindent The speedup can be measured, we know the number of parallelized threads (or processing units),
so $\alpha$ provides an 'empirical parallelization'.
As we can easily conclude from the simple model of parallelized sequential processing (see Fig.~1), $\alpha$
actually corresponds to the ratio of \textit{the time} of the payload computations to the \textit{total} measurement time.
One can visualize Amdahl's assumption that (assuming many processors)
in $\alpha=Payload/Total$ fraction of the measured processing time, the processors are processing data,
in (1-$\alpha$) fraction they are waiting (all but one). That is $\alpha$ describes
how much, on average, processors are utilized.

This quantity defines the empirical factor that includes
everything, from the engineering perfectness  of
assembling the components to the sequential-only fragments
of the computation to the delay of the internal connections.
According to Amdahl, \textit{any fraction that we cannot parallelize,
contributes to the sequential-only fraction}.
As Amdahl discussed, and can quickly be concluded from Fig.~1, the sequential-only
fraction has contributions from the software, from the
operating system, from the hardware. Even, the science also provides its contribution: the physical
size of the computer plays a role. At extreme sizes,
the speed of signal propagation (and of course the technology components
of the network) lead to a considerable increase of the 'idle time'.
In many-processor systems serious
competition takes place for the shared resources (such as the buses), the queuing of the messages inside the system
seriously increases the sequential-only fraction, and so
it considerably reduces the parallelizable fraction
of the processing time.
Depending on the actual situation, those contributions have a very different order of magnitude, and also they can compete
for dominance.

\subsection{The performance of parallel processing}
The fact that the parallel efficiency has its limitations
was recognized early, and even its reason was correctly identified:
"\textit{As pointed out by Amdahl~\cite{AmdahlSingleProcessor67}, the [constant
	problem size] CPS scaling leads to
	a rapid reduction in parallel efficiency as more processors are used to solve a fixed-size, deterministic problem. It was Amdahl argued that most parallel programs have
	some portion of their execution that is inherently serial
	and must be executed by a single processor while others
	remain idle."}~\cite{ScalingParallel:1993}

\begin{figure}
	\includegraphics[scale=1.1]{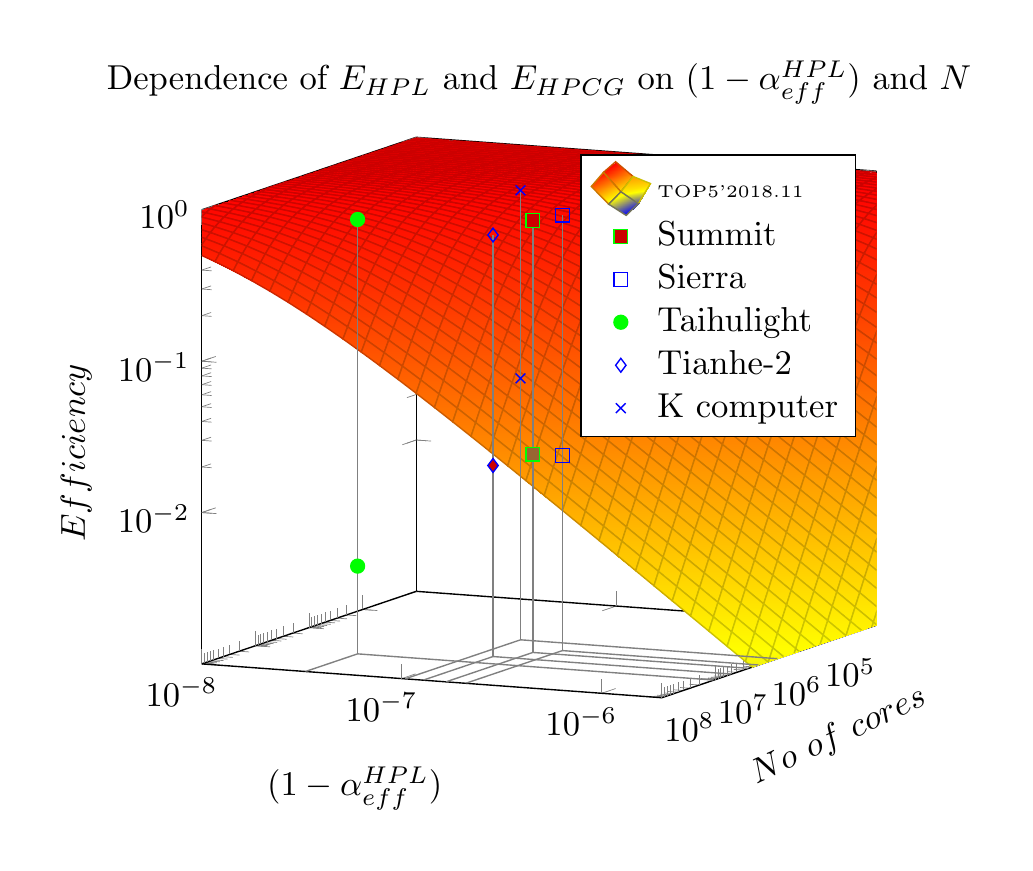}

	\caption{The efficiency of the performance of parallelized computing systems on the parameters number of cores and efficiency of parallelization  The data points for the present top supercomputers are derived using data in the TOP500 database. The top marks on the vertical lines refer to benchmark $HPL$, the bottom marks to $HPCG$.	\label{fig:efficiency2dim}}
\end{figure}

When calculating speedup, one actually calculates
\begin{equation}
S=\frac{(1-\alpha)+\alpha}{(1-\alpha)+\alpha/N} =\frac{N}{N(1-\alpha)+\alpha}
\end{equation}
hence  the \textit{efficiency}\footnote{This quantity is almost exclusively used to describe computing performance of multi-processor systems.
	In the case of supercomputers, $\frac{R_{Max}}{R_{Peak}}$ is provided, which is identical with $E$, excepts that it is actually a 2-parameter function, see Fig.~2 and the bottom row of Table~I.} (how speedup scales with the number of processors)
\begin{equation}
E = \frac{S}{N}=\frac{1}{N(1-\alpha)+\alpha}\label{eq:soverk}
\end{equation}

\noindent hence  the \textit{efficiency}
\vspace{-.3\baselineskip}
\begin{equation}
\boxed{E(\large N,\alpha)} = \frac{S}{N}=\boxed{\frac{1}{\textcolor{webred}{\Large N}(1-\alpha)+\alpha}}= \frac{R_{Max}}{R_{Peak}} \label{eq:soverk}
\end{equation}

\noindent That is, according to Amdahl, the efficiency depends  on two critical
variables: parallelization efficiency (the mathematical details are discussed in~\textbf{\cite{Vegh:2017:AlphaEff}}) and the number of the processing units (or threads), as shown in Fig.~2%\ref{fig:efficiency2dim}
. The surface corresponds to the algorithmic dependence
of the efficiency on those two variables\footnote{Recall that
	"\textit{this decay in performance is not a fault of the architecture,
		but is dictated by the limited parallelism}."\cite{ScalingParallel:1993}.}. The data points of efficiency
shown at different values of the scaled parameters
at the given numbers of respective scaling parameters
for some of the TOP25 supercomputers.

The inefficiency was attributed (at that time correctly) to the software contribution
"\textit{for example, include
	initialization, task creation, or some other phase of computation}". The key sentences that the processors spend
"\textit{some portion of their execution that is inherently serial and must be executed by a single processor while others remain idle}" and "\textit{scaling thus put larger machines at an inherent disadvantage}"~\cite{ScalingParallel:1993} remained out of sight for a considerable while.
As Fig.~2 %\ref{fig:efficiency2dim}
witnesses, $Taihulight$, and $K~computer$ stand out from the "million core" group. Thanks to its 0.3M cores, $K~computer$ has the best efficiency for the $HPCG$ benchmark, while $Taihulight$ with its 10M cores the lowest one (the good efficiency for $HPL$ originates
in using "cooperating processors"~~\cite{CooperativeComputing2015}).
The middle group follows the rules. For $HPL$ benchmark: the more cores, the lower efficiency. For $HPCG$ benchmark:
the "roofline"~\textbf{\cite{VeghRoofline:2019}} of that communication intensity reached,
they are about the same efficiency.

\subsection{The syndrome of 'idle processors'}
At that time both the parallel efficiency of the hardware
and the number of processors was low.
However, while in 1993 the
TOP500 list one can find "supercomputers" assembled from only 2 (or even 1) processors,
the clock frequency was about 1\% of the today's clock frequency
and the perfectness of their parallelization~\textbf{\cite{Vegh:2017:AlphaEff}} was about
$10^{-3}$, for today, the top supercomputers comprise millions of processors, and their perfectness of parallelization is about $10^{-7}$.
That is, all important factors have changed by orders of magnitude.
Today, despite the huge improvement of parallel efficiency
and the respectable single-processor performance,
the large number of processors dominates. \textit{Millions of processors are
	idle --because of both software and hardware reasons--
	while one processor executes the serial-only fraction of the code}~\textbf{\cite{VeghPerformanceWall:2019}}.
In contrast with~\cite{ScalingParallel:1993}, that "\textit{the serial fraction \dots is a diminishing function of the problem size}", the weight of the housekeeping activity grows linearly with the number of the cores (and so does the idle time of the other cores). The issue known since decades returned in a technically different form at a vast number of cores.
After reaching a critical number of processors (using the
present paradigm and technology, one can guess its value to be slightly less than 10M), \textbf{\textit{adding more processors leads to a decreasing performance}}~\cite{VeghPerformanceWall:2019,ScalingParallel:1993,NeuralScaling2017}, as witnessed by demonstrative supercomputer failures such as  $Gyoukou$, $Aurora$ or $SpiNNaker$. It was also noticed that a larger number of computing nodes, the performance starts to decrease~\cite{ToolingUpForExascale:2019}. The mathematical details are discussed in~\textbf{\cite{VeghPerformanceWall:2019,VeghBrainAmdahl:2019}}.

At the time of writing the paper~\cite{ScalingParallel:1993}, the quantitative breakdown of the factors of inefficiency were not yet sufficiently known:
the relatively inefficient HW parallel efficiency
(as well as the high contribution of the software)
did not enable us to study the other contributions.
Unfortunately, later different other scalings were also
introduced (sometimes not with the proper care, see~\cite{UsesAbusesAmdahl:2001}), \textit{suggesting that the
	parallel efficiency can be enhanced infinitely, i.e.
	that the computing performance of parallel systems has no (at least close-lying) limit}.

Some researchers, however, correctly guessed that initialization, threading,  are the main reasons of inefficiency.
The software also contributes: the different need for synchronization of the different tasks causes different supercomputer efficiencies~\cite{DifferentBenchmarks:2017}
and, of course, the payload computation also needs time.
All this leads to different "rooflines" of parallelized computing~\textbf{\cite{VeghRoofline:2019}}. As otherwise also predicted by~\cite{ScalingParallel:1993}, the amount of synchronization (data and control communication) of the processors is a decisive factor in defining the efficiency~\cite{CriticalSectionAmdahlEyerman:2010,SynchronizationEverything2013,YavitsMulticoreAmdahl2014}. Another critical factor can be a  "higher-level synchronization" such as the commonly used "biological clock time" in brain simulation~\textbf{\cite{VeghBrainAmdahl:2019}} that also decreases the parallel efficacy of the computing systems by orders of magnitude.
This paper discusses how much the parallel efficiency depends on those different factors.

\section{The efficacy of parallel processing\label{sec:ParallelEfficacy}}

The architectures based on the classic paradigm,
the '\textit{\textbf{S}ingle \textbf{P}rocessor \textbf{A}pproach}', have serious limitations~\cite{AmdahlSingleProcessor67}; among others their efficiency strongly degrades~\cite{ScalingParallel:1993} as the number of parallelized units increases.
The quick development of the underlying technology
(the fake promise of the Moore-observation) covered
the need to look for ways of utilizing 'cooperating processors', as advised.
For today, it became obvious that the single-processor performance shall reach its limits within years because of some laws of nature~\cite{LimitsOfLimits2014}, but today the computing is prepared for the post-Moore era~\cite{Post-Moore_2017}, rather than looking for a new computing paradigm. Under the term of "new computing paradigm", using gates other than those built from transistors
even when thinking about rebooting computing~\cite{RebootingComputingModels:2019} or speculations on utilizing
computing in some specific utilization fields, is understood.

\subsection{The role of the communication}
As was analyzed a quarter of century before us~\cite{ScalingParallel:1993}, the principle of the parallelization itself creates an additional bottleneck~\cite{InherentSequentiality:2012}~\textbf{\cite{VeghPerformanceWall:2019}}. As seen also from our model, at the beginning and end of the measurement, most of the processors are idle.
The bottleneck is the non-parallelizable fraction of the job: according to Amdahl, this factor alone limits the achievable speedup (or performance gain: the resulting performance compared to the performance of a component processor).
The researchers also recognized that \textit{one of the major contributors to the bottleneck is the needed communication} between the parallelized sequential units.
We also know that at a given technology of parallelization, after some point, the computing performance of the system not only saturates, but (because of the speedily growing fraction of housekeeping) \textit{starts to decrease when increasing further the number of processors in the parallelized system}~\cite{ScalingParallel:1993}.
At that time, the critical number was a few dozens of processors,
see Fig.~1 in~\cite{ScalingParallel:1993}; today, one can guess it to be at a few million processing units.

\subsection{Why the issue re-appeared}

The technology of parallelization at that time was at the stage~\cite{GordonBellPrize:2017}
that the performance gain
(despite that thousands of processors were used in the systems) could not achieve 200. According to that stage,
one considered the major bottleneck to be the software contribution
(like initialization, thread creation), at that time correctly. It was stated~\cite{ScalingParallel:1993} that
"\textit{the serial fraction, s, in most applications does not remain constant but is a diminishing function of the problem size, Amdahl’s law does not apply directly when the problem is scaled}". The benchmarking time in the case of supercomputers is in the order of several hours~\cite{DongarraSunwaySystem:2016} (or $10^{13}$ clock cycles), so the non-parallelizable fraction (the one-time initialization and task creation) becomes small, but remains finite.
That small sequential-only fraction, while one processor organizes the work (and talks individually to all fellow processors), shall be multiplied by the total number of processors in the system\footnote{One can decrease this number by clustering or more successfully using 'internal clustering'~\cite{CooperativeComputing2015}}, so the formerly 'diminishing part' is amplified strongly. Not
only is it non-negligible, but can even dominate. For a detailed discussion see~\textbf{\cite{VeghPerformanceWall:2019}}.

Although that statement was valid at the time of writing paper~\cite{ScalingParallel:1993}, in the age of several dozens of processors only,
it must be revised when there are millions of processors in the computing systems. Nowadays, the time when one processor is performing the non-parallelizable activity while the others
remain idle, shall be multiplied by a factor of $10^3$\dots$10^6$ times higher (so much more times idle time).

During the past quarter of a century, the number of processors increased, and the efficiency of their parallelization enhanced.
Now it is time to scrutinize the roles of the different contributions. As it turned out~\textbf{\cite{VeghPerformanceWall:2019}},
the technical development reordered ranking of the different
contributions to the non-parallelizable fraction of computing. In the modern systems, the HW parallelization efficiency (mainly due to the higher interconnection speed) can be smaller than that of the already mentioned software contribution. Due to the larger sizes, one must re-evaluate the role of the physical size and the interconnection.

For today,
\textit{the need of communication between the parallelized parts became a dominant factor} ~\cite{ScalingParallel:1993}. It is well is known that at large core numbers, the efficiency of the supercomputers also depends on the task running on the supercomputer~\cite{DifferentBenchmarks:2017}.
As experienced for the two popular benchmark programs HPL and HPCG, there are as many efficiencies as many benchmarks.
The lack of understanding of the role of the number of processors and the parallelization efficiency leads to explanations that some 'architectural weakness' is behind the much lower efficiency of \textit{Taihulight} compared with the \textit{K-computer}.
	The real reason is the 30-times higher number of the processors and Amdahl's law. And, of course: the efficiency depends on the task, more precisely on the amount of communication the task requires.

\begin{figure}
	\includegraphics{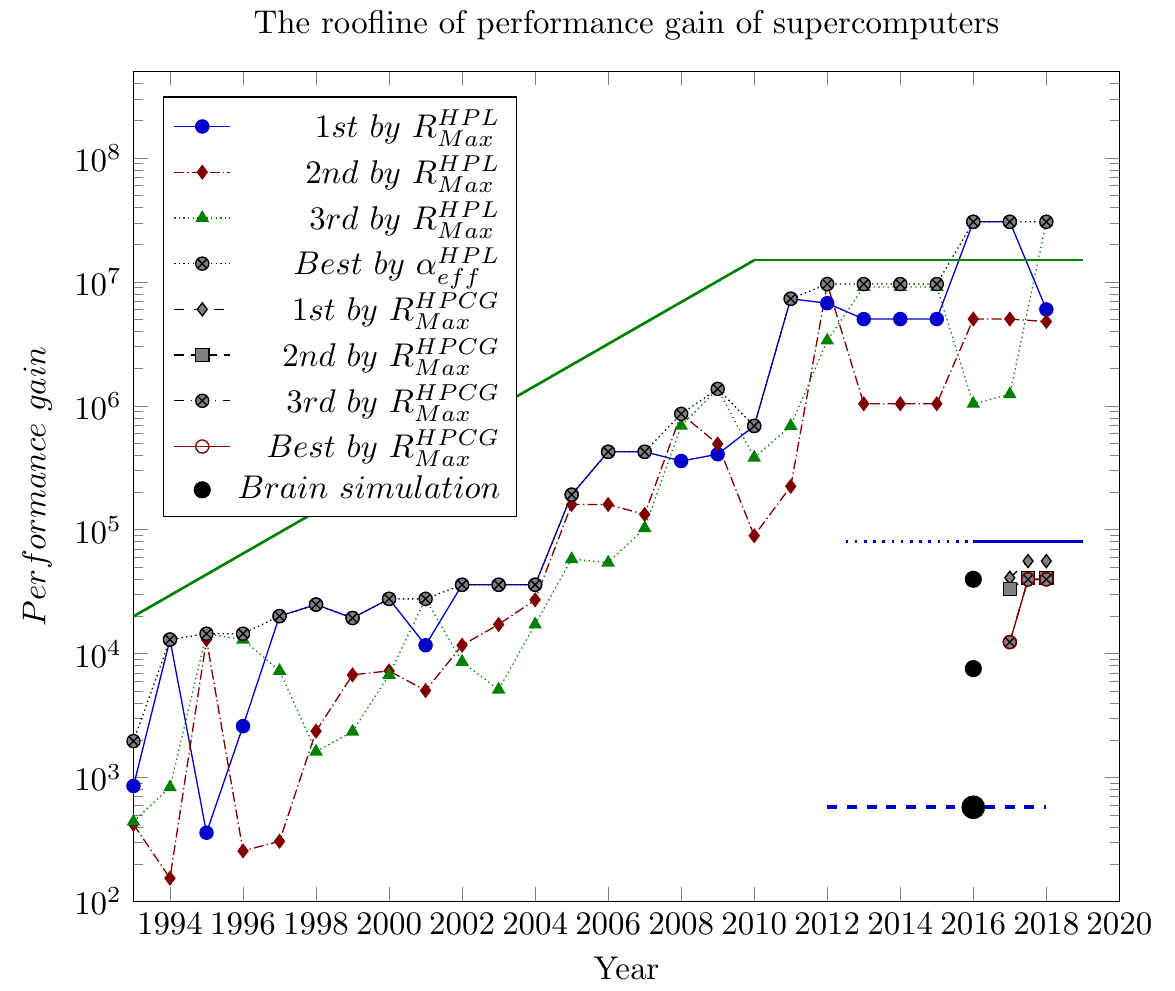}

	\caption{The rooflines of the performance gain, for different levels of communication: the minimum level (the HPL benchmark), the medium level (the HPCG benchmark) and the very high level (brain simulation)~\protect{\cite{VeghBrainAmdahl:2019}} \label	{fig:rooflines}}
\end{figure}

As the careful analysis enabled to conclude~\textbf{\cite{VeghRoofline:2019}}, the mentioned
'inherently non-parallelizable contribution'~\cite{InherentSequentiality:2012}~\textbf{\cite{VeghPerformanceWall:2019}} of the task
effectively represents a resource that limits the achievable performance, so the popular 'roofline model'~\cite{WilliamsRoofline:2009}
can also be applied~\textbf{\cite{VeghRoofline:2019}} to the achievable performance of the parallelized systems, see Fig.~3.%
%\ref{fig:rooflines}.

\section{The need for 'modern computing'\label{sec:ModernComputing}}

The case of computing is very much analogous with the case of classic physics versus the modern (relativistic and quantum) physics~\textbf{\cite{VeghModernParadigm:2019}}.
In the world we live in, it is rather counter-intuitive to accept
that as we move towards unusual conditions, the adding of speeds behaves differently,
the energy becomes discontinuous, the momentum and the position of a particle cannot be measured accurately at the same time.
The computations result in no reasonable difference in the cases we experience around us, with and without using the non-classical principles. However, as we get farther from the normal conditions, the difference gets more considerable, and even leads to phenomena one can never experience under the usual, everyday conditions. \textit{The analogies do not want to derive direct correspondence between certain physical and computing phenomena. Rather, this paper wants to call the attention to both that under extreme conditions, qualitatively different behavior may be encountered, and that scrutinizing certain, formerly unnoticed or neglected aspects enables to explain the new phenomena. In computing, unlike in nature, the technical implementation of the critical points can be changed. In this way the behavior of the computing systems can also be changed.}
In this paper, only the affected (and for AI operation relevant) important area of computing can be touched in more detail: parallel processing.

\subsection{Anomalies in extreme-scale computing
\label{sec:Anomalies}}
In both previous figures, one can see an anomaly (the same anomaly from two different points of view).

In Fig.~2 %\ref{fig:efficiency2dim}
the only 10M processor supercomputer $Taihuight$
stands out from the group of other top-class supercomputers, and also in Fig.~3 %\ref{fig:rooflines}
the HPL performance gain values for $Taihuight$ fall above the corresponding roofline level. The reason is the same: their processor~\cite{CooperativeComputing2015} utilizes direct core-to-core data transfer, i.e., the computing principle is (slightly) different.
The amount of communicated data furthermore the computations are the same as that at the other supercomputers,
but the communication takes place without using the global bus systems; because of this, the communication takes less time.

Reducing the communication makes sense.
The so-called HPL-AI benchmark uses Mixed Precision
\cite{MixedPrecisionHPL:2018} rather than Double Precision
computations. This change in the operand length enabled them to achieve nearly 3 times better
performance gain, that (as correctly stated in the announcement)
"\textit{Achieving a 445 petaflops mixed-precision result on HPL (equivalent to our 148.6 petaflops DP result)}", i.e. the peak DP performance did not change.
Unfortunately, this achievement has not much to do with
AI: it utilizes the data representation commonly used in AI,
but \textit{the achievement comes from accessing less data in memory and using quicker operations
	on the  shorter data
	rather than reducing the communication intensity}.
For AI applications, the limitations remain the same
as described above;
except that when using Mixed Precision,
the efficiency can be better by a factor of 2-3.
For other effects, see also section~\ref{sec:efficacyAI}.

This performance increase clearly shows that not the number of communication operations to computation operations, but the $time$ spent with those operations defines the efficiency of the parallel system. Thanks to this, $Taihuight$ is the only competitor with 10M cores; all the others have only a fragment of this number. This difference is the reason is why $Gyoukou$ disappeared in a half year after it was able to utilize only 12~\% of its 20M available cores, and this is why $Aurora$ did not appear.
The phenomena with large-scale systems show that computing under extreme conditions --such as large scale parallel systems-- behaves differently compared to just a few processor systems.

\begin{table}
	\begin{tabular}{|p{180pt}|p{180pt}|}
		\hline
		\hline
		Physics & Computing\\
		\hline
		Adding of speeds &	Adding of performance\\
		\hline
		\textcolor{blue}{Classic} & \textcolor{blue}{Classic} \\
		\textcolor{blue}{$ v(t) = t\cdot g$}
		&
		\textcolor{blue}{$ Perf_{total}(n) = n\cdot Perf_{single}$}	\\
		\hline

		c = Light Speed &\\
		\hline
		t = time &  n = number of cores\\
		\hline
		g = acceleration & $Perf_{single}$\\
		\hline
		n = optical density
		&
		$\alpha$ = parallelism\\
		\hline
		\textcolor{red}{ Modern (relativistic)} &\textcolor{red}{Modern\cite{VeghAlphaEff:2016}} \\
		\textcolor{red}{$ v(t) = \frac{t\cdot g}{\sqrt{1+{(\frac{t\times g}{c/n}})^2}}$}
		&
		\textcolor{red}{$ Perf_{total}(n) = \frac{n\cdot Perf_{single}}{n\cdot (1-\alpha)+\alpha}$}
		\\
		\hline
		\hline

	\end{tabular}
	\vskip\baselineskip
	\caption{The "adding speeds" analogies between the classic and modern arts of science and computing, respectively	\label{tab:summing}}

\end{table}

The supercomputers are stretched to the limit~\cite{StrechingSupercomputers:2017,Scienceexascale:2018}.
It underlines the importance of the engineering perfectness that
in the case of Summit, adding 5~\% more cores (and making fine-tuning after its quick startup) resulted in 17~\% increase in the computing performance in a half year after its appearance on the list, and in another half year a 0.7~\% increase in the number of cores caused a 3.5~\% increase in its performance.
The one-time appearance of $Gyoukou$ is mystic:
it could catch slot \#4 on the list using just 12\% of it 20M cores,
although the explicit ambition was to be the \#1.
The lack of understanding that under extreme conditions,
the computing performance behaves differently led to the false presumptions of frauds when reporting computing performances ~\cite{GyoukouFraud:2017}.
The another champion candidate, supercomputer Aurora~\cite{DOEAuroraMistery:2017} --after years of building--
was retargeted, weeks before its planned startup.
In November 2017   Intel announced
that \textit{Aurora} has been shifted to 2021. As part of the announcement, the development line \textit{Knights Hill}~\cite{IntelDumpsXeonPhi:2017} was canceled, and instead be replaced by a "new platform and new microarchitecture specifically designed for exascale". The lesson learned was that \textit{one needs specific design for exascale}.

\subsection{Analogy with the special relativity
\label{sec:AnalogyRelativity}}

In the above sense, there is an important difference between the operation and the performance of the single-processor
and those of the parallelized but sequentially working computer systems.
As long as just a few (thousands) single processors are aggregated
into a large computer system, the resulting performance corresponds (approximately) to the sum
of the single-processor performance values: similarly to the classic rule of adding speeds.
However, when assembling larger computing systems (and approaching with their performance "the speed of light" of computing systems in the range of millions of processors)  the experienced \textit{payload} performance
starts to deviate from the \textit{nominal} performance:
the phenomenon known as \textit{efficiency} appears.

The performance measurements are simple time measurements: the benchmark program executes a standardized set of machine instructions
(a large number of times), and the known number of operations is divided by the measurement time.
This procedure happens in the same way in the case of measuring the performance,
for the single-processor and the parallelized sequential computing systems.
In the latter case, however, the joint work must also be organized. With that activity, an extra task (implemented with additional machine instructions and additional execution time) appears; see also Fig.~\ref{fig:AmdahlModelAnnotated}.
 Due to this, \textit{the computing performance cannot increase above the performance defined by (the parallelization technology and) that number of processors},
in analogy with that an object having the speed of light cannot be further accelerated.
\textit{Exceeding a specific computing performance (using the classic paradigm and its implementation) is prohibited by the laws of nature}.

\begin{figure}
	\hspace{-1.5cm}
	\includegraphics[scale=1.15]{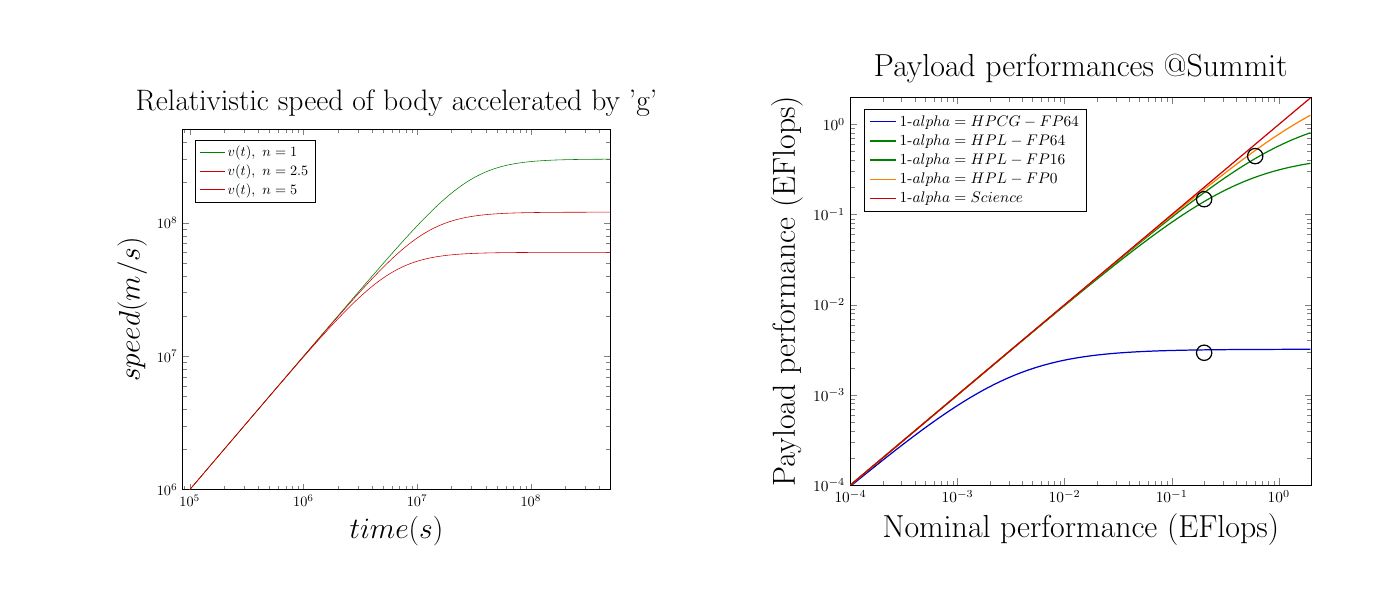}
	\vspace{-2\baselineskip}
	\caption{The effect of the correction term for the relativistic acceleration and the payload performance
		of supercomputers.\label{fig:SummitLimits}}
\end{figure}

Table~\ref{tab:summing} and Fig.~\ref{fig:SummitLimits}
show why the analogy is relevant.  Although their function forms
are different, the modern approach introduces in both cases a correction term. That term remains close to unity until the extreme conditions approached, then
both functions saturate: the specific "Light Speed" cannot be exceeded.
The figure shows the performance of the current "world champion" Summit. The diagram lines show the
performance calculated from the theory, for the cases of
(from bottom-up) the HPCG benchmark, the HPL benchmark (both with FP-64),
the HPL-AI (FP-16) benchmark~\cite{MixedPrecisionHPL:2018}.
The next diagram line shows the estimated performance for the FP-0 case, that is when no floating operations are executed,
but the program executes non-floating operations like addressing, comparing, incrementing, jumping.
(this latter is similar to the "empty loop", frequently used in programming.)
The top diagram line corresponds to the case when only
the physical size of the computer limits the performance
(but the computer makes nothing useful).
The bubbles refer to the values measured on the system
Summit. As shown, at low $nominal$ performances,
the $payload$ values of performance show up no considerable difference.
At higher performances, however, the \textit{payload performance depends on what task the supercomputer runs}.

\subsection{Analogy with the general relativity
\label{sec:AnalogyGeneralRelativity}}

The mentioned losses manifest in the appearance of the performance wall~\textbf{\cite{VeghPerformanceWall:2019,VeghRoofline:2019}},
a new limitation due to the parallelized sequential computing.
In science, we know that the enormous masses behave differently from what we know under 'normal' conditions.
If we assume we know how the 'matter' behaves, we need to assume
the presence of 'dark matter'. The latter is like the 'matter' but not quite. Which is another phrase to tell that the large scale behavior of 'matter' largely deviates from that we have concluded from the smaller amount of 'matter'. An analogy with this field is in use already: the phenomenon of 'dark silicon'~\cite{Computing_Dark_Silicon_2017} is named in analogy with the 'dark matter':
the (silicon) cores are there, and usable,
but (because of the thermal dissipation) the large amount of cores behaves differently.

Analogously, the parallel computing introduces the "dark performance":
the cores are there and clocked, they consume power, but they make
no payload work: they are idle.
Because of the principle of the classic computing,
the first core must speak to all fellow cores,
and \textit{this non-parallelizable fraction of the
	time increases with the
	number of the cores}~\textbf{\cite{VeghRoofline:2019,VeghPerformanceBook:2019}}. The result is that the top supercomputers (depending on the number of cores) show up efficacy around 1\% when solving real-life tasks. In analogy with the "gravitational collapse", a "communicational collapse" is demonstrated in Fig.~5.(a) in ~\cite{CommunicationCollapse:2018}: at an extremely large number of cores exceeding the critical threshold of communication intensity leads to unexpected and drastic change of network latency.

The propagation time of signals is also very much similar to that of the effects of physical fields.  The latency time of the interfaces can be paired with
creating and attenuating the physical carriers. Zero-time on/off signals are possible both in the classical physics and classical computing,
while in the corresponding modern counterparts also the time needed to create, transfer and detect the signals must be accounted; the effect noticed as that the time of wiring (in this extended sense) grows compared to the time of gating~\cite{LimitsOfLimits2014}.

\subsection{Analogy with the quantum physics
\label{sec:AnalogyQuantum}}
The electronic computers are clock-driven systems, i.e., no action can happen in a time period shorter than the length of one clock period.   The typical value of that "quantum of time" in computing today is in the nanosecond range, so in the everyday praxis, the time seems to be continuous, and the "quantal nature of time" cannot be noticed.  Some (sequential) non-payload fraction in the total time is always present in the parallelized sequential systems,
not only because of spawning and joining the threads but also because of their concurrency~\cite{InherentSequentiality:2012}. That fraction cannot be smaller than the ratio of the length of two clock periods divided by the total measurement time, since also the forking and joining the other threads cannot be shorter than one clock period.
Unfortunately, the technical implementation needs about ten thousand times longer time to do those actions~\cite{Tsafrir:2007,armContextSwitching:2007}.

The total time of the performance measurement is large (typically hours) but finite, so the non-parallelizable fraction is small but finite.
Because of Amdahl's Law (the computing paradigm and the
implementation technology together), the absolute value of \textit{the computing performance of parallelized systems has inherently an upper limit, and the efficiency is the lower, the higher is the number of the aggregated processing units}.

As discussed below and in detail in ~\textbf{\cite{VeghBrainAmdahl:2019}},
the processor-based brain simulation provides an
"experimental evidence" that the time in computing shows quantal behavior, analogously with the energy in physics. When simulating neurons utilizing processors,
the ratio of the simulated (biological) time and the processor time used to simulate the biological effect may considerably differ, so to avoid working with "signals from the future", periodic synchronization is required that introduces a particular "biological clock cycle".
The role of this clock period is the same as that of the clock signal in the clocked digital electronics: what happens in this period, it happens "at the same time".

The commonly used $1~ms$~\cite{NeuralNetworkPerformance:2018} "grid time" is,
however, $10^6$ times longer than the
$1~ns$ clock cycle common in the digital electronics.
Correspondingly, its influence on the performance is noticeable,
see the subfigure Fig.~\ref{fig:alphacontributions}.C.
As shown, the "quantal nature of time" in computing
changes the behavior of the performance drastically.
Not only the achievable performance is by orders of magnitude lower,
but also the "communicational collapse" (see also~\cite{ScalingParallel:1993}) occurs at orders of
magnitude lower nominal performance.
This effect is the reason why less than one percent of the planned capacity
can be achieved even by the purpose-built brain simulator~\cite{SpiNNaker:2013} as well as that the SW based and HW based simulations show up the same limitation~\cite{NeuralNetworkPerformance:2018,VeghBrainAmdahl:2019}.
The memory of huge supercomputers can be populated~\cite{SpikingPetascale2014} with objects simulating neurons, but as soon as they need to start to communicate,
the task collapses as predicted in Fig.~\ref{fig:alphacontributions}. This reasoning is indirectly underpinned~\cite{NeuralScaling2017} by that the different handling
of the threads changes the efficacy sensitively and that the time required for more detailed simulation increases non-linearly~\cite{NeuralNetworkPerformance:2018,VeghBrainAmdahl:2019}.

\subsection{Analogy with the interactions of particles
\label{sec:AnalogyInteraction}}
The ability of \textit{communicating} with each other is not a native feature of processors in the 'classic computing':
in the \textit{\textbf{S}ingle \textbf{P}rocessor \textbf{A}pproach} questions like message sending to and receiving from some other party
has no sense at all (as no other party exists); messaging is very ineffectively imitated by SW in the layer between HW and the real SW.
In the \gls{SPA} the
communication is a non-parallelizable fraction of the activity of the cores,
and similarly sharing resources has no sense (although it is an elementary requirement in all modern systems).

The laws of parallel computing result in the actual behavior of the computing systems:
the more communication takes place~\textbf{\cite{VeghRoofline:2019}},
the more deviation from the classic behavior is experienced. Similarly,
as in physics, the behavior of an atom sharply changes by the interaction (communication) with other particles.

\subsection{Rebooting our thinking
\label{sec:RebootingThinking}}

"Classic computing" cannot explain these phenomena.
The limits of single-processor performance enforced by the laws of nature~\cite{LimitsOfLimits2014} are topped by the limitation of parallel computing~\textbf{\cite{VeghPerformanceWall:2019,VeghRoofline:2019}}.
The "quantal nature of time"~\textbf{\cite{VeghBrainAmdahl:2019}} limits that performance further.
Notice that these contributions are competing with each other;
the actual circumstances decide which of them can dominate.
Their effect, however, is very similar: according to Amdahl, \textit{what is not parallel is qualified as sequential}.
To understand the phenomena, we introduced a modern computing paradigm~\textbf{\cite{VeghModernParadigm:2019}} (as opposed to the 70-year old classic paradigm).

Anyhow, if we want to completely understand the
phenomena experienced under extreme (computing) conditions,
we must reboot also our thinking; using another technology
implementation as "rebooting computing models"~\cite{RebootingComputingModels:2019} is not sufficient.
Today we have extremely inexpensive (and at the same time: extremely complex and
powerful) processors around (a "free resource"~\cite{SpiNNaker:2013}) and we arrived to the age when no additional reasonable
functionality can be implemented in processors through adding more transistors, the over-engineered processors optimized for single-processor regime do not enable reducing the clock period~\cite{EPIC:2000}.
The computing power hidden in many-core processors cannot be utilized effectively
for payload work, because of the "power wall" (\textit{partly because of the improper working regime}~\cite{EnergyProportional2007}):
we arrived at the age of "dark
silicon"~\cite{Computing_Dark_Silicon_2017},
we have "too many" processors~\cite{TooManyCores2007} around.
The supercomputers face critical efficiency and performance issues;
the real-time (especially the cyber-physical) systems experience serious predictability,
latency and throughput issues;  in summary, the computing performance (without
changing the present paradigm) reached its technological bounds.
Computing needs renewal~\textbf{\cite{RenewingComputingVegh:2018}}.
Our proposal, the \textit{Explicitly Many-Processor Approach (EMPA)}~\textbf{\cite{IntroducingEMPA2018}},
is \textit{to introduce a new computing paradigm} and through that \textit{to reshape the way in which computers (including ANNs) are designed and used} today.
\begin{figure}
	\maxsizebox{\textwidth}{\textheight}
	{
		\begin{tabular}{cc}
			\raisebox{30ex}{\huge HPL}&	\includegraphics{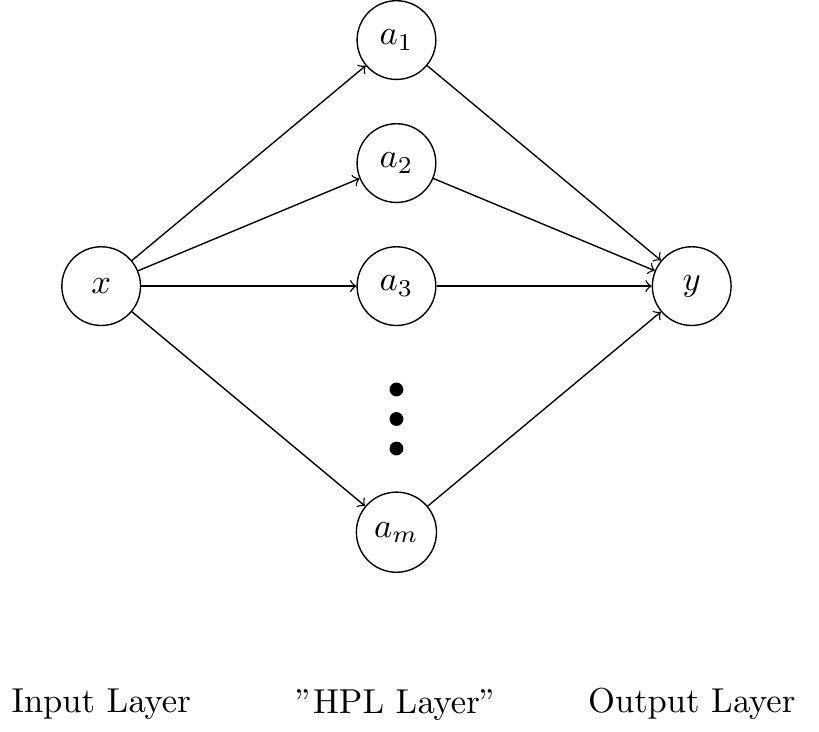}
			\\
			\raisebox{30ex}{\huge HPCG} &	\includegraphics{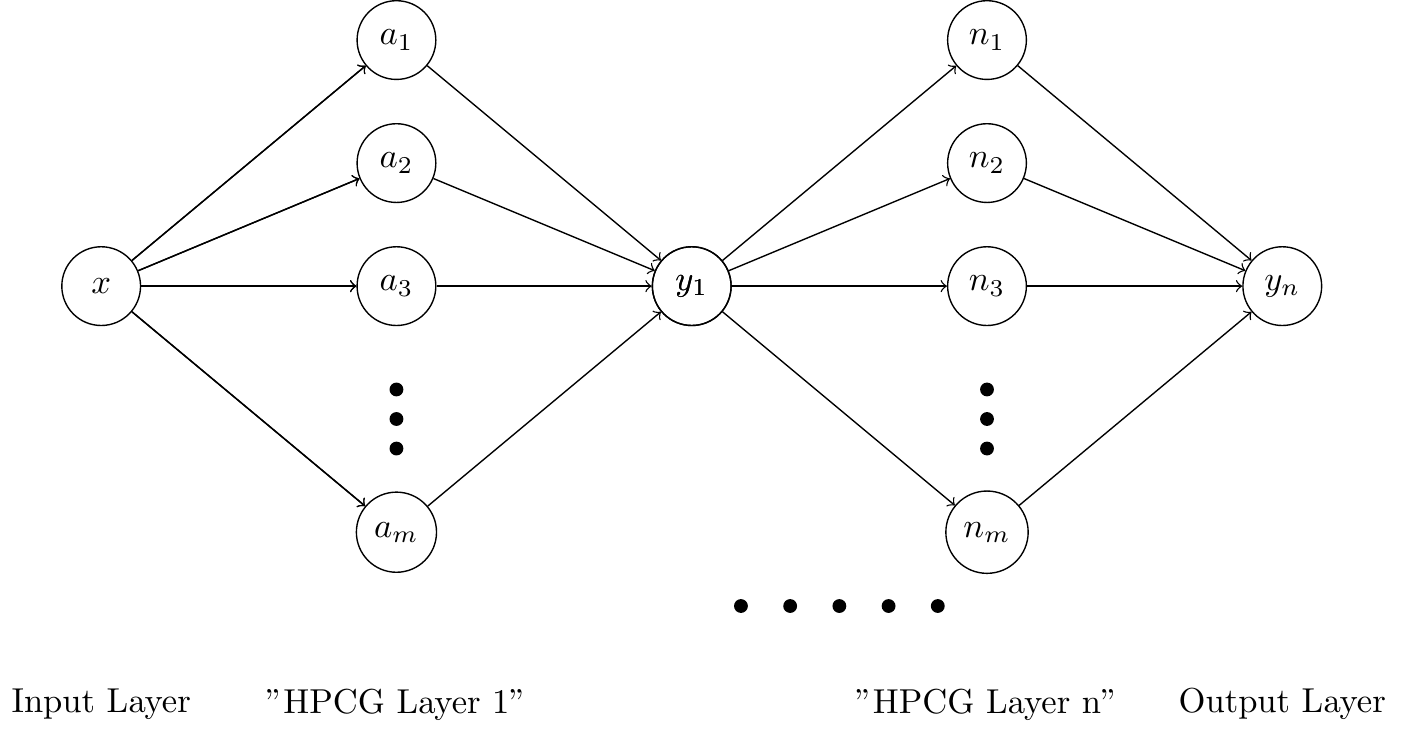}
			\\
			\raisebox{30ex}{\huge AI} &	\includegraphics{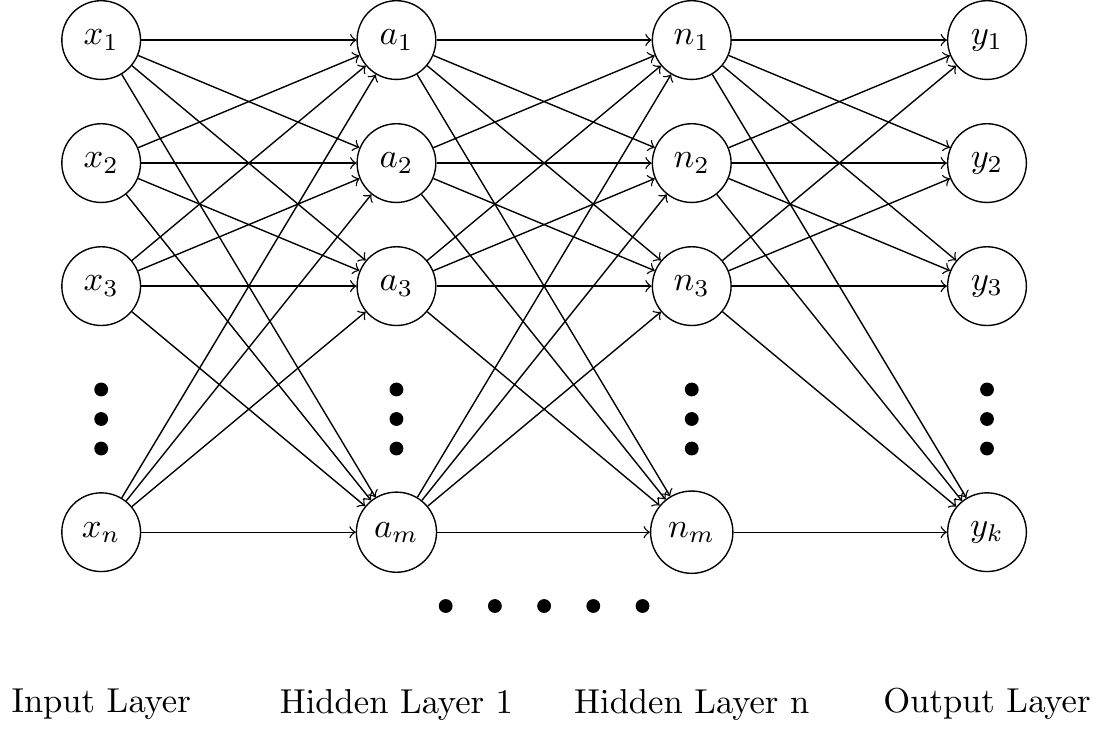}
		\end{tabular}
	}
	\caption{The communication intensity in the different supercomputer tasks.
		The subfigures correspond to the HPL, HPCG and AI cases.
		The arrows represent a communication action.	\label{fig:multilayers}}

\end{figure}

\section{The effect of synchronization on the efficiency\label{sec:synchronizationAI}}

The engineering practice commonly uses the synchronization of the different and independently working circuits. On one side, it makes the conditions clear:
the time dependence of the signals is removed in this way. On the other side,
however, it synchronizes also the needs of communication: the different units
want to send their result at the same time and also they receive their input at the same time;
leading to a considerable amount of idle waiting.

\subsection{Benchmarking and efficiency}
Communication is a dominating factor in the operation of the many-many processor systems.
Fig.~\ref{fig:multilayers} attempts to provide a feeling of how the different tasks behave for different communication intensity. The communication to computational intensity~\cite{ScalingParallel:1993} is, of course, not proportional in the cases of the subfigures, but the figure illustrates excellently how the communication need of the different computer tasks changes with the type of the task.

There are two commonly used benchmarks in supercomputing.
As discussed in~\textbf{\cite{VeghRoofline:2019}}, the HPL class tasks essentially need communication only at
the very beginning and at the very end of the task. This communication type is, however, not the way as the real-life programs work. Because of this difference, the benchmark HPCG was introduced in a couple of years ago: the experience shows that the expectable payload performance is much more accurately approximated by HPCG than by HPL, because the real-life tasks need much more communication.
The supercomputers show different efficiencies when using different benchmark programs~\cite{DifferentBenchmarks:2017}. The efficiencies differ by a factor of ca. 200-500, when measured by HPL and HPCG, respectively, mainly due to the differing number of cores.

In Fig.~\ref{fig:multilayers} three special cases (owing
different communication intensity) are compared.
In the top and middle figures the benchmark communication
intensities of the popular supercomputer benchmarks $HPL$ and $HPCG$ are displayed in the style of AI networks.
The "input layer" and "output layer" are the same,
and comprise the initiating node only, while the
other "layers" are again the same: the rest of the cores.
Subfigure 6.C depicts an AI network comprising
$n$ input nodes and $k$ output nodes, furthermore
$h$ hidden layers comprising $m$ nodes.

In the HPL class, the communication intensity is the lowest possible one:
the computing units receive their task (and parameters) at the beginning
of the computation, and they return their result at the very end.
That is, the core coordinating their work must deal with the fellow cores only in these periods, so the communication intensity is proportional to the
number of the cores in the system. Notice the need to queue the requests at the beginning and the end of the task.

In HPCG class, iteration takes place: the cores return the result of one iteration to the coordinator core, which makes sequential operations: not only receives and re-sends the parameters, but also needs to compute the new parameters before sending them to the fellow cores, and repeats this several times. As a consequence, the non-parallelizable fraction of the benchmarking time grows proportionally with the number of iterations.
The effect of the extra communication decreases the achievable performance roofline: as shown in Fig.~\ref{fig:rooflines},
the HPCG roofline is about 200 times lower than the HPL one.
The fact that supercomputers show two different efficacy for these two different benchmarks~\cite{DifferentBenchmarks:2017} is well known, but its reason is not fully understood. Rather than comprehending that the reason is simply the effect of the number of cores on the efficiency of the parallelized sequential systems, some "architectural weaknesses" are supposed.

As can be easily seen from the figure,
in the case of the benchmark $HPL$ the initiating node must
issue $m$ communication messages and collect $m$ returned results,
i.e. the execution time is $O(2m)$.
In the case of the benchmark $HPCG$ this execution time is
$O(2Nm)$ where $N$ is the number of iterations.
(One cannot compare the execution times directly because of the
different amount of computations).

\subsection{Supercomputer efficiency in terms of AI}
The bottom part of Fig.~\ref{fig:multilayers} depicts how the Artificial Neural Networks
are supposed to operate. The life begins in several input channels (rather than one
as in the HPL and HPCG cases) that would be advantageous.
However, the values must be communicated to \textit{all} nodes in the top hidden layer:
the more input nodes and the more nodes in the hidden layer(s), the many $times$ more communication is required for the operation. The same also happens when the
first hidden layer communicates data to the second one, except that here \textit{the square of the number of the nodes} is to be used as a weight factor of communication.

Initially the $n$ input nodes
issue messages, each one $m$ messages (queuing\#1) to the nodes
in the first hidden layer, i.e., altogether $nm$ messages.
If one uses a shared bus to transfer the messages, these  $nm$ messages must be queued (queuing\#2).
Also, every single node in the hidden layer
receives (and processes) $m$ input messages (queuing\#3).
Between the hidden layers, the same is repeated (maybe several times)
with $mm$ messages, and finally $km$ messages
are sent to the output nodes.
In all cases, the messages are queuing 3 times.
To make a fair comparison with benchmarks $HPL$ and $HPCG$,
let us assume one input and one output node.
In this case, the AI execution time is $O(h\times m^2)$,
provided that $h$ hidden layers are implemented.
(Here it was assumed that the messaging mechanism
between layers is independent from each other.
It is not so if they share a global bus.
\footnote{\textit{"The idea of using the popular shared bus to implement the communication medium is no longer acceptable, mainly due to its high contention."}~\cite{ReconfigurableAdaptive2016}})

For a numerical example: let us assume that
in the supercomputers 1M cores are used, and
in the AI network, 1K nodes are present in the hidden layers,
and only one input and output nodes are used.
In that case, all execution times are $O(1M)$
(again, the amount of computation is sharply different,
so the scaling can be compared, but not the execution times).
This communication intensity explains why in Fig.~\ref{fig:rooflines}
the $HPCG$ "roofline" falls
hundreds of times lower than that of the $HPL$:
the increased communication need strongly decreases
the achievable performance gain.

Notice that the number \textit{computation} operations increases with $m$,
while the number of \textit{communication} operations with $m^2$. In other words:
the more nodes in the hidden layers, the higher is the communication intensity (communication/computation ratio), and because of this, the lower is the efficiency of the system.
Recall that since the AI nodes perform simple computations
compared to the functionality of the supercomputer benchmarks,
the communication/computation ratio is much higher,
making the efficacy even worse.
The conclusions are underpinned by experimental research~\cite{DeepNeuralNetworkTraining:2016}:
	\begin{itemize}
	\item "strong scaling is stalling after only a few dozen nodes"
	\item "The
	scalability stalls when the compute times drop below the communication
	times, leaving compute units idle. Hence becoming a communication bound
	problem."
	\item "the network layout
	has a large impact on the crucial communication/computation
	ratio: shallow networks with many neurons per layer \dots scale worse than deep networks with less neurons."
\end{itemize}

The massively "bursty" nature of the data (the different nodes of the layer
want to use the communication at the same moment) also makes the case harder.
The commonly used global bus is overloaded with messages.
The possibility for wired point-to-point communication is limited;
but deploying them at least for the inter-layer communication buses can help a lot.

The communication circuits receive the task of sending the data to $N$ other nodes.
The computation and communication are \textit{ab ovo} sequential, and
the communication channel can only transfer one data value at a time.
What is worse, bus arbitration, addressing, latency, prolong the transfer time (and in his way decreases the efficacy of the system).

\subsection{The case of brain simulation\label{sec:brainsimulation}}
The case of brain simulation is greatly similar to the combined HPCG and AI case.
The number of layer nodes have fellow neurons in the range $10^3$\dots$10^4$,
i.e., after one computational step the layer neurons must wait up to
$10^4$ times the communication time. When assuming that the communication time
is $10^2$ times longer than the computation time, it means that the efficacy of such a system can be about $10^5$\dots$10^6$ times
lower than the efficacy without this type of communication. Even if one does NOT consider another limiting (technical) factors, like the bandwidth of communication.

One one side, even with today's large number of available cores in the top supercomputers,
several orders of magnitude are missing to achieve the units in the brain.
On the other side, a processor core today has much more computing performance
than needed to simulate the operation of a single neuron.
The need and possibility leads to resource sharing: the computing capacity of
one core is shared by several 'neurons'. To rearrange the scene (to switch to another thread) needs anyhow extra
organizational  (again, sequential-only) time, and the context switching is extremely expensive~\cite{Tsafrir:2007,armContextSwitching:2007}.

The brain simulation (and in somewhat smaller scale: artificial neural computing) requires intensive data exchange between the parallel threads:
the neurons are expected to tell the result of their neural computations periodically to thousands of fellow neurons.
Because the neurons must work on the same (biological) time scale,
the (commonly used)  1~millisecond "grid time" ("the quantum of computing time") has a noticeable effect on the performance.

As pointed out in~\textbf{\cite{VeghBrainAmdahl:2019}}, using the commonly accepted 1~ms
'time slot'
(the 'biological clock period' or the manifestation of the
'quantal nature of time') alone degrades the efficacy of the simulator program, which is topped
by the large amount of burst-like communication. As a consequence, the maximum
performance gain of supercomputer programs (or purpose-built, but processor based
nervous simulators) cannot  exceed $\approx10^4$, although the idea of introducing
at least partly hierarchic communication paths like\cite{IntelLoihi:2018}
can reduce by about 2 orders of magnitude the communication\footnote{Loihi is a purpose-built computing chip, i.e., its performance on its
	special field, as demonstrated, can be about hundreds of times higher
	than that of the competitors, but  the resulting scaling in extreme size systems
	also undergoes limitations.}, i.e., it can increase the performance gain by about 2 orders of magnitude.
The efficacy of the AI networks must be between the efficacy of the real-life supercomputer tasks and that of the brain simulation; for details see~\cite{VeghRoofline:2019}.

\section{Efficacy of AI networks\label{sec:efficacyAI}}

In the history of computing, the "need for speed" is a central question. The different efforts to develop the single-processor
performance wanted to achieve a higher number of operations,
literally at any price. The case of parallelized processing is not different. Since the limits of single-processor performance are approached, and also the limitations of the parallel operation are not entirely discovered, one must be careful with the new designs. To avoid some traps:
to make some features better, one must make some other worse.

As mentioned above, to achieve high performance \textit{and} low latency at the same time is not possible, so the proper balance between them depends heavily on the conditions of the application area.

\subsection{Factors affecting efficacy of the systems}
\begin{figure}
	\includegraphics[scale=.9]{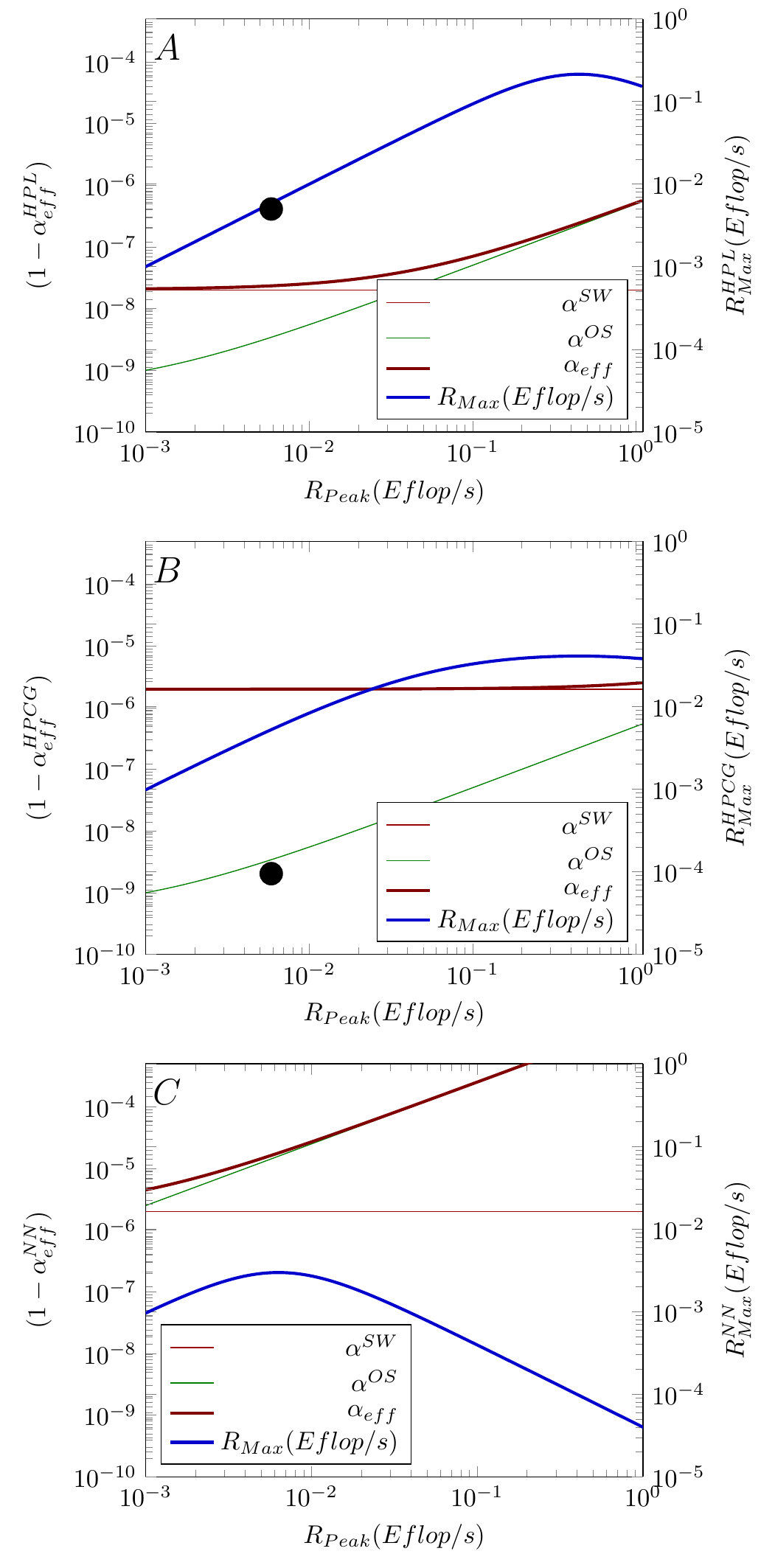}
	\caption{Contributions $(1-\alpha_{eff}^X)$ to $(1-\alpha_{eff}^{total})$ and max payload performance $R_{Max}$ of a fictive supercomputer ($P=1Gflop/s$ @ $1GHz$) in function of the nominal performance.
		The blue diagram line refers to the right hand scale ($R_{Max}$ values), all others ($(1-\alpha_{eff}^{X})$ contributions) to the left scale. The figure is purely illustrating the concepts; the displayed numbers are somewhat similar to the real ones.\label{fig:alphacontributions}
	}

\end{figure}

Fig.~\ref{fig:alphacontributions} attempts to summarize the role of the different contributions to the resulting efficacy of parallelized systems.
The subfigure A illustrates the behavior measured with benchmark \gls{HPL}. The looping contribution becomes remarkable around 0.1~Eflops, and breaks down payload performance when approaching 1~Eflops. The black dot marks the \gls{HPL} performance of the computer used in works~\cite{NeuralNetworkPerformance:2018,NeuralScaling2017}.
Subfigure B depicts the behavior measured with benchmark \gls{HPCG}. In this case, the contribution of the application (thin brown line) is much higher, the looping contribution (thin green line) is the same as above. As a consequence, the achievable payload performance is lower, and also the breakdown of the performance is softer.
The black dot marks the \gls{HPCG} performance of the same computer.
Subfigure C demonstrates what happens if the clock cycle is 5000 times longer: it causes a drastic decrease in the achievable performance and firmly shifts the performance breakdown toward lower nominal performance values.

Subfigures $A$ and $B$ roughly correspond to the supercomputer benchmark cases ($HPL$ and $HPCG$), respectively, the subfigure $C$ approximately
illustrates the case of AI performance. As shown,
using the present principles and implementations,
the performance of parallelized sequential systems is
drastically lower than expected from the nominal performances.
The actual values depend on the actual methods of implementation,
i.e., the methods and principles of the implementation (including both the architecture and the mathematical algorithm)
of the AI networks should be carefully selected
if one wants to achieve not only the functionality,
but also a proper efficacy.

\subsection{The role of accelerators}
As an illustration, the special role of some accelerators is discussed here.
"\textit{How those accelerators connect to systems is a significant concern}"~\cite{OpenCAPI:2019}, really.
On one side, they help to perform the computations more quickly.
On the other side, they increase the latency, i.e. decrease efficacy.
The communication intensity is a decisive factor in the operation of the system,
both in latency and efficacy.
The communication intensity can be enhanced either by increasing the proportion spent
with computations or decreasing the time of data communication (increasing the communication bandwidth or using other methods of communication).

\begin{figure}
	\includegraphics[scale=1.15]{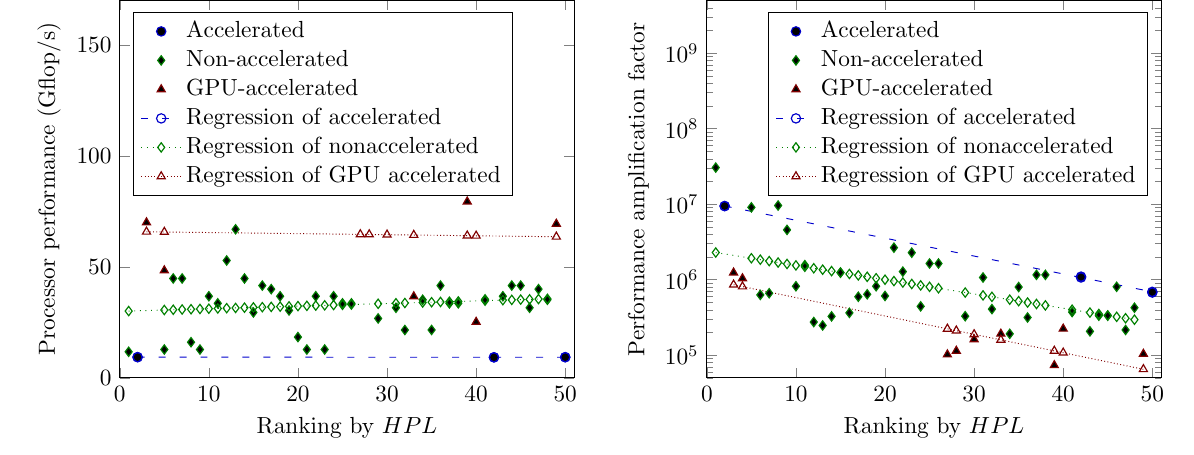}
	\caption{The effect of the GPGPU acceleration on
	the single-processor performance and parallelization efficiency. The data are taken from the database~\cite{Top500:2016}.\label{fig:GPGPUcorrelation}}

\end{figure}

Using GPGPUs to accelerate the computing is very popular also in supercomputing, and the huge and rigorously controlled database~\cite{Top500:2016}
enables us to demonstrate these two effects.
Fig.~7 depicts the dependence of the "single processor performance" and "parallelization efficiency" for supercomputers with and without GPGPU accelerators, in the function of ranking in 2017.
As the figure shows, the GPGPU accelerated \textit{performance} --independently of the ranking-- is about
2-3 times higher than that of the non-accelerated one. The right subfigure displays that the \textit{performance gain} changes with ranking.\footnote{This is due to the increased latency caused by the need to copy the data from one memory to another. Introducing OpenCAPI~\cite{OpenCAPI:2019}  enhances the latency (see the success of Summit and Sierra), but presently no statistically evaluable data are available on that.}
At higher ranking (at a lower number of cores)
the performance amplification is higher than 4,
while at a high number of cores (at low ranking)
the performance amplification is only slightly above 2. This result is in good accordance with the result of a systematic CPU/GPU performance comparison~\cite{Lee:GPUvsCPU2010}.

One must, however, consider that in the case of the AI networks the network latency times are about a hundred times higher than the computational times, so making the $computations$ faster does not decrease the bottleneck.
On one side, the absolute time of processing decreases. On the other side, the communication intensity (i.e., the efficiency) gets worse.
The effect is very much similar to that of the processor accelerator GPGPUs: the computing performance increases \textit{linearly}, the efficiency decreases \textit{exponentially}.
These accelerators behave differently on small scale and large scale systems.
On small scale systems, only the performance increase can be noticed; the number of the cores is low. On high scale systems, the large number of cores increases the latency (i.e., decreases the efficacy). That is, new principles for data communication must be sought~\cite{ReconfigurableAdaptive2016}.

\subsection{Traps of enhancing systems's efficacy}

Preparing systems with high payload computing performance is a more complex task than commonly assumed.
The efficacy changes with the number of processing units and the contributions to the non-parallelizable portion of the task $\alpha$. Both of them must be concerted to achieve high performance.  It is not something new, again:
"\textit{%A fairly obvious conclusion which can be drawn at this point is that
the effort expended on achieving high parallel processing rates is wasted unless it is accompanied by achievements
in sequential processing rates of very nearly the same magnitude}"~\cite{AmdahlSingleProcessor67}.

On one side, the demonstrative failures of supercomputers with an extremely large number of processing units prove that
using the conventional architecture (the conventional contributions of the technical components) the number of processing
units are limited to about 1 million.
On the other side, it is shown that the clustering
successfully attacks the looping contribution,
although it was already noticed that at extremely large number of nodes a similar performance breakdown occurs~\cite{ToolingUpForExascale:2019}.
The chip-internal clustering introduced by~\cite{CooperativeComputing2015} is successful, at least
enables us to operate 10+ million processing units.
That solution can produce outstanding computing performance when measured via HPL benchmark, but
for real-life tasks (having efficacy similar to that of the
HPCG benchmark) Amdahl' law limits the performance. The reason for that difference is that the dominating
contribution changes between the two benchmarks: in HPL
the contributions from interconnection+HW+OS dominate~\textbf{\cite{VeghModernParadigm:2019}},
while in HPCG the dominating contribution comes from
the computation+communication (i.e., the "measuring device" itself).

The GPGPU acceleration, as discussed, also behaves differently
in small-scale and large-scale systems, see Fig.~7. The basic reason (without using OpenCAPI~\cite{OpenCAPI:2019} interconnection)
is that the data must be copied between the different
address spaces that increases the sequential-only
component.

Ironically enough, using lower precision also hides a
similar trap. As discussed, the communication/computation ratio
is one of the decisive factors of the payload performance.
When using half-precision instead of double-precision,
the computation time shall be four times less, but the
communication time remains the same, i.e., the communication to computation ratio shall be higher. On one side, the absolute amount of time spent with computation+communication decreases,
but so does the total measured time, too. As a consequence,
the contribution to the (Relative!) factor $(1-\alpha)$ is
less than expected. When using half-precision arithmetics,
Summit consumes four times less power~\cite{MixedPrecisionHPL:2018}, but the execution time
of the benchmark HPL is only three times shorter.
Consequently, \textit{using shorter floating operands~\cite{MixedPrecisionHPL:2018} and new representations of floating numbers~\cite{GustafsonFloating:2018} in AI applications  increase the relative communication intensity}, and through this decrease the efficiency (although it is counter-balanced
by the decreased computing time).
Besides, the reduced precision also has its limitations~\cite{HalfPrecisionArithmetic:2017}.

\begin{figure}
	\maxsizebox{1.05\textwidth}{!}
	{
		\hspace{-1cm}
		\begin{tabular}{cc}
			\includegraphics[scale=1.1]{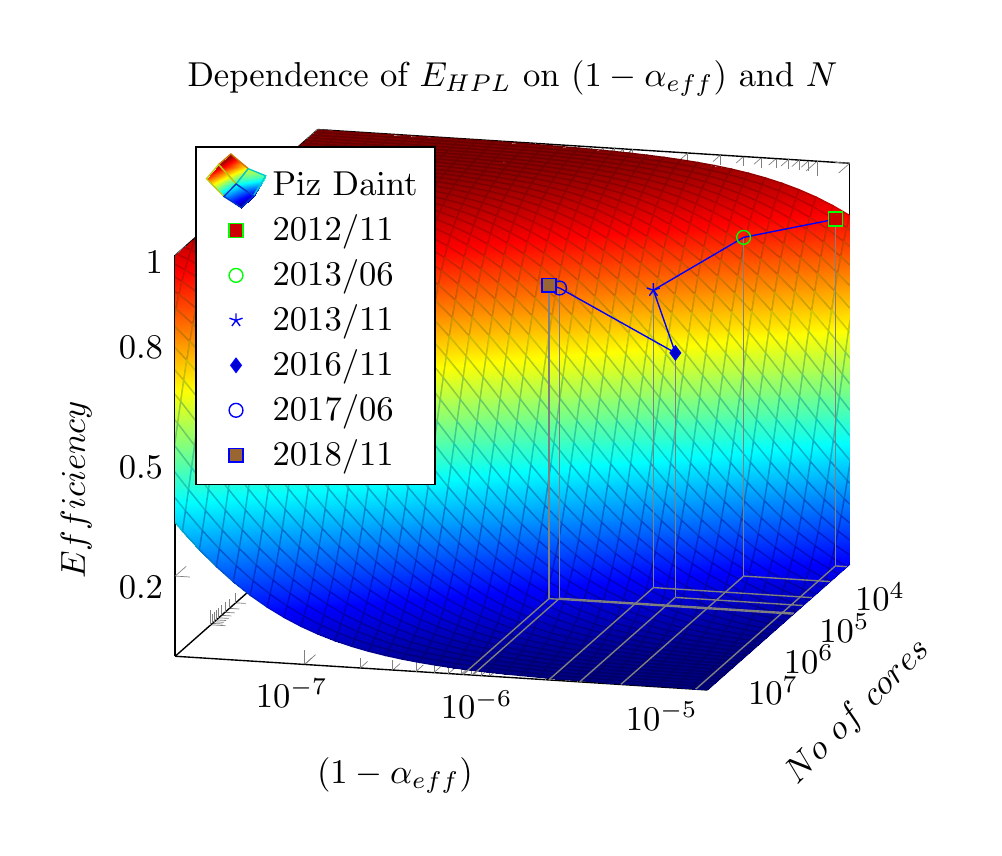}
			&
			\includegraphics[scale=.785]{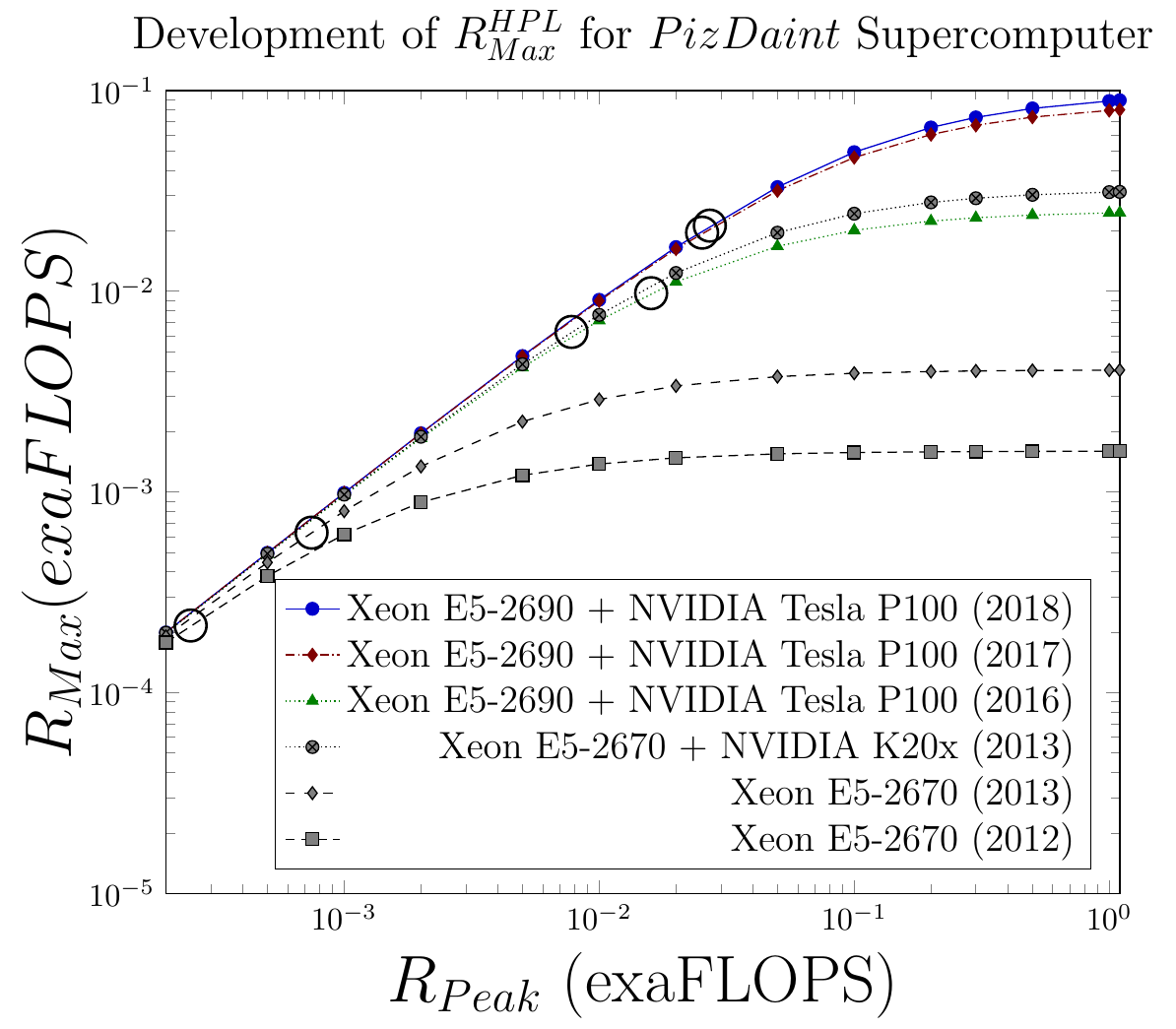}
		\end{tabular}
	}
	\caption{The history of supercomputer Piz Daint in terms of  efficiency and  payload performance~\cite{Top500:2016}.\label{fig:PizDaintEfficiency}}

\end{figure}

\subsection{The case of PizDaint}
The supercomputers usually have not many items registered in the database TOP500  on their development. One of the rare exceptions is supercomputer $Piz~Daint$. Its development history spans 6 years, two orders of magnitude in performance, and used both non-accelerated computing and accelerated computing using two different accelerators.
Although usually more than one of its parameters was changed between the registered stages of its development,
it nicely underpins the statements of this paper.
Fig.~\ref{fig:PizDaintEfficiency} displays how the \textit{payload performance} in the function of the \textit{nominal performance} has developed in the case of supercomputer $Piz~Daint$, and how at the same time the
efficacy evolved.

In the right subfigure, the bubbles display the measured performance values documented in the database TOP500~\cite{Top500:2016}, and the diagram lines show the (at that stage) predicted performance. As the diagram lines show the "predicted performance", the accuracy of the prediction can also be estimated through the data measured in the next stage. The left subfigure shows the efficacy values
on the 2-parameter efficacy surface, in the same stages
of development.

The data from the first two years of $Piz~Daint$ (non-accelerated mode of operation)
can be compared directly. Increasing the number of the cores results in the expected higher performance, as the working point is still in the linear region of the efficiency surface. The value slightly above the predicted one can be attributed to the fine-tuning of the architecture.

Introducing accelerators resulted in a jump of payload efficiency (and also moved the working point to the slightly non-linear region, see Fig.~\ref{fig:PizDaintEfficiency}), and the payload performance is roughly 3 times more than it would be expected purely
on the predicted value computed from the non-accelerated architecture. According to the general experience~\cite{Lee:GPUvsCPU2010}, \textit{only a small fraction of the computing power hidden in the \textit{GPU} can be turned to payload performance}.

The designers might not be satisfied with the accelerator, so they changed to another one, with a slightly higher nominal performance but much larger separated memory space. The result was disappointing:
the slight increase of the nominal performance of the \textit{GPU} could not counterbalance
the increased time needed to copy between the separated larger address spaces, and finally resulted in a breakdown
of both the value of $(1-\alpha_{eff})$ and efficiency, although the payload performance slightly increased.
Introducing the \textit{GPU} accelerator increases the absolute performance,
but (through introducing the extra non-parallelizable component of copying the data) increases the value of $(1-\alpha_{eff})$ and decreases efficiency.

\section{Conclusion}\label{sec:summaryAI}

The biology-inspired computing (such as Artificial Neural Networks, Deep Learning, Brain Simulation) are exciting
and useful fields. One must not forget, however, that
there are fundamental differences between the biologically and electronically implemented systems.
The clock-driven, parallelized sequential electronic systems, in some cases, can only very ineffectively imitate the behavior of the biological systems. The paper wanted to
call attention to some key efficacy questions of their design.

\section*{Acknowledgements}
Project no. 125547  has been implemented with the support provided from the National Research, Development and Innovation Fund of Hungary, financed under the K funding scheme.

%\bibliographystyle{spphys}
%\bibliography{../../CommonBibliography,%
%	../../CommonPrivateBibliography%
%}

%++------------------------------------------------------------------------------

\label{lastpage-01}

\end{document}